%% file: main_postprint.tex
\documentclass[10pt,conference,comsocconf]{IEEEtran}
\IEEEoverridecommandlockouts

\usepackage{cite}
\usepackage{amsmath,amssymb,amsfonts}
\usepackage{algorithm}
\usepackage{algpseudocode}
\usepackage{graphicx}
\usepackage{textcomp}
\usepackage{xcolor}
\def\BibTeX{{\rm B\kern-.05em{\sc i\kern-.025em b}\kern-.08em
    T\kern-.1667em\lower.7ex\hbox{E}\kern-.125emX}}

\usepackage[acronym]{glossaries}
\usepackage{subfigure}
\usepackage{multirow}
\usepackage{booktabs}

\newlength{\subfigsep}
\usepackage{makecell}

\usepackage{fancyhdr}

\usepackage{multirow}

\usepackage{tcolorbox}

\usepackage{tabularx}

\begin{document}


\thispagestyle{empty} 

\vfill 

\begin{center}

\begin{tcolorbox}[
    colback=gray!10, colframe=black, fonttitle=\bfseries,
    width=0.9\textwidth, boxrule=1pt, arc=5pt, outer arc=5pt,
    boxsep=10pt, left=10pt, right=10pt, top=10pt, bottom=10pt
]
\textbf{THIS IS AN AUTHOR-CREATED POSTPRINT VERSION.}

\vspace{0.3cm}

\textbf{Disclaimer:}  
This work has been accepted for publication in the \textit{Joint European Conference on Networks and Communications \& 6G Summit (EuCNC/6G Summit)}, 2026.  

\vspace{0.3cm}

\textbf{Copyright:}  
© 2026 IEEE. Personal use of this material is permitted. Permission from IEEE must be obtained for all other uses, in any current or future media, including reprinting/republishing this material for advertising or promotional purposes, creating new collective works, for resale or redistribution to servers or lists, or reuse of any copyrighted component of this work in other works.

\vspace{0.3cm}


\end{tcolorbox}
\end{center}

\vfill 

\clearpage
\setcounter{page}{1}

\input{acronyms.tex}

\title{Performance Analysis of 5G RAN Slicing Deployment Options in Industry 4.0 Factories}

\author{\IEEEauthorblockN{Oscar Adamuz-Hinojosa, Abdelhilah Abdeselam, Pablo Muñoz, Pablo Ameigeiras, Juan M. Lopez-Soler}

\IEEEauthorblockA{Department of Signal Theory, Telematics and Communications, University of Granada. \\ Email: abdelhilah@correo.ugr.es,  \{oadamuz,pabloml,pameigeiras,juanma\}@ugr.es}
}

\maketitle

\begin{abstract}
This paper studies Radio Access Network (RAN) slicing strategies for 5G Industry~4.0 networks with ultra-reliable low-latency communication (uRLLC) requirements. We compare four RAN slicing deployment options that differ in slice sharing and in the degree of per-line or per-flow isolation. Unlike prior works that assume a fixed slicing structure, this work addresses how RAN slicing should be instantiated in the presence of multiple production lines and heterogeneous industrial flows. A Stochastic Network Calculus (SNC)-based analytical framework and a heuristic slice planner are used to evaluate per-flow delay guarantees and radio resource utilization. Within the considered RAN-level analytical model, the results show that, under resource scarcity, only per-flow slicing prevents delay violations, whereas slice-sharing and hybrid deployments improve aggregation efficiency at the cost of weaker protection for the most delay-critical flows. Execution-time results show that the proposed planner operates at non-real-time (Non-RT) time scales, supporting its implementation as an \textit{rApp} within Open RAN (O-RAN) Non-RT RAN Intelligent Controller (RIC) control loops.
\end{abstract}


\section{Introduction}
Recent advances in \gls{5G} are accelerating the adoption of wireless communications in Industry~4.0, enabling \gls{IIOT}, cyber-physical systems, and \gls{RT} automation~\cite{Wollschlaeger2017}. Critical industrial applications require deterministic \gls{uRLLC} performance to ensure the reliable operation of production lines, controllers, and sensors. Despite \gls{5G} \gls{uRLLC} features such as flexible numerologies, mini-slots, and grant-free access~\cite{Le2021}, stringent latency and reliability guarantees remain difficult under wireless channel variability and heterogeneous traffic. Under congestion, heavily loaded flows may degrade others, directly affecting production accuracy in functions such as motion control, controller synchronization, and actuation. This creates a clear need for performance isolation under scarce radio resources.

Network slicing has emerged as a key enabler of logical networks with tailored \gls{QoS} requirements~\cite{Adamuz2023}. By dedicating radio resources, slices can provide latency and reliability guarantees while isolating production-line traffic, making them particularly relevant for smart factories~\cite{5GACIA_QoS_2021}. However, most existing works focus on architectures or resource allocation algorithms without systematically addressing \gls{RAN} slicing-level deployment. When multiple production lines generate heterogeneous flows with diverse \gls{QoS} requirements, the impact of \gls{RAN}-level slicing decisions on isolation, delay guarantees, and resource efficiency remains insufficiently understood.

\textbf{Literature Review}. Prior studies on \gls{RAN} slicing for industrial scenarios mainly address architectural frameworks, slice management, or resource allocation mechanisms. Rashid et al.~\cite{Rashid2022} proposed slice descriptors to enhance reliability, while Perdigão et al.~\cite{Perdigao2024Automating} studied automated slice management for factory environments. Other works investigated flexible slicing for heterogeneous traffic~\cite{Zhang2022} or analyzed slicing architectures in edge federation and robotic systems~\cite{Taleb2019,Mhatre2021}. Complementary studies examined the limitations of current \gls{5G} releases for industrial traffic~\cite{Lucas-Estañ1} and spectrum-level slicing strategies with their trade-offs~\cite{Munoz2020}, while experimental platforms such as 5G-CLARITY demonstrated the feasibility of infrastructure-level slicing and multi-connectivity in private industrial networks~\cite{Cogalan2022}. However, these works generally assume a predefined slicing structure and do not systematically characterize alternative \gls{RAN} slicing deployment options within a single factory, nor their impact on isolation, per-flow delay guarantees, and radio resource utilization.

\textbf{Contributions}. The above discussion raises the following question: \emph{how should \gls{RAN} slicing be instantiated in a factory to support heterogeneous \gls{QoS} requirements across multiple industrial flows?} To address it, this paper makes three contributions: \textit{C1}) identification and comparative analysis of multiple \gls{RAN} slicing deployment options in industrial factories, clarifying their impact on isolation and per-flow delay guarantees; \textit{C2}) a \gls{SNC}-based analytical framework to derive per-flow delay bounds, used as a common basis for consistent comparison across deployment options; and \textit{C3}) a heuristic slice planner that allocates downlink radio resources to satisfy per-flow delay targets.

\textbf{Paper Organization}. Section~\ref{sec:SlicingStrategies} introduces the slicing options. Section~\ref{sec:SystemModel} presents the traffic, resource, and channel models. Section~\ref{sec:EstimationSNC} details the \gls{SNC}-based planner. Section~\ref{sec:perf_results} reports the evaluation results. Section~\ref{sec:conclusions} concludes the paper.

\section{Network Slicing Deployment Options}\label{sec:SlicingStrategies}

\begin{figure*}[t!]
  \centering
  \begin{minipage}[t]{\dimexpr0.5\textwidth-0.5\subfigsep\relax}
    \centering
    \includegraphics[width=0.95\linewidth]{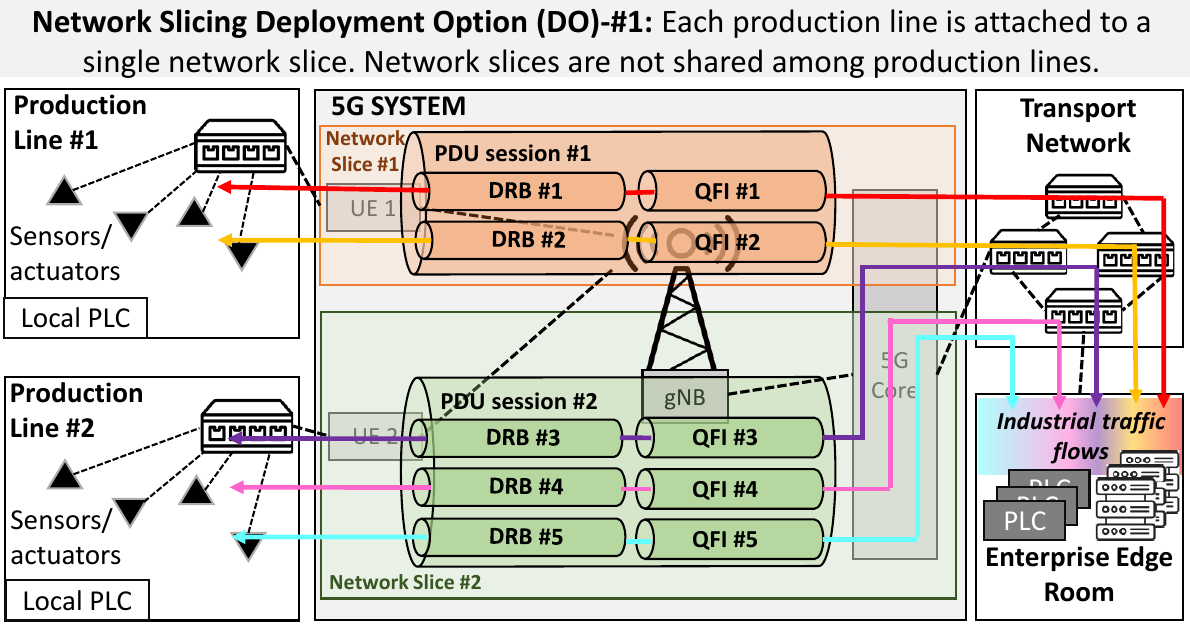}
  \end{minipage}\hspace{\subfigsep}%
  \begin{minipage}[t]{\dimexpr0.5\textwidth-0.5\subfigsep\relax}
    \centering
    \includegraphics[width=0.95\linewidth]{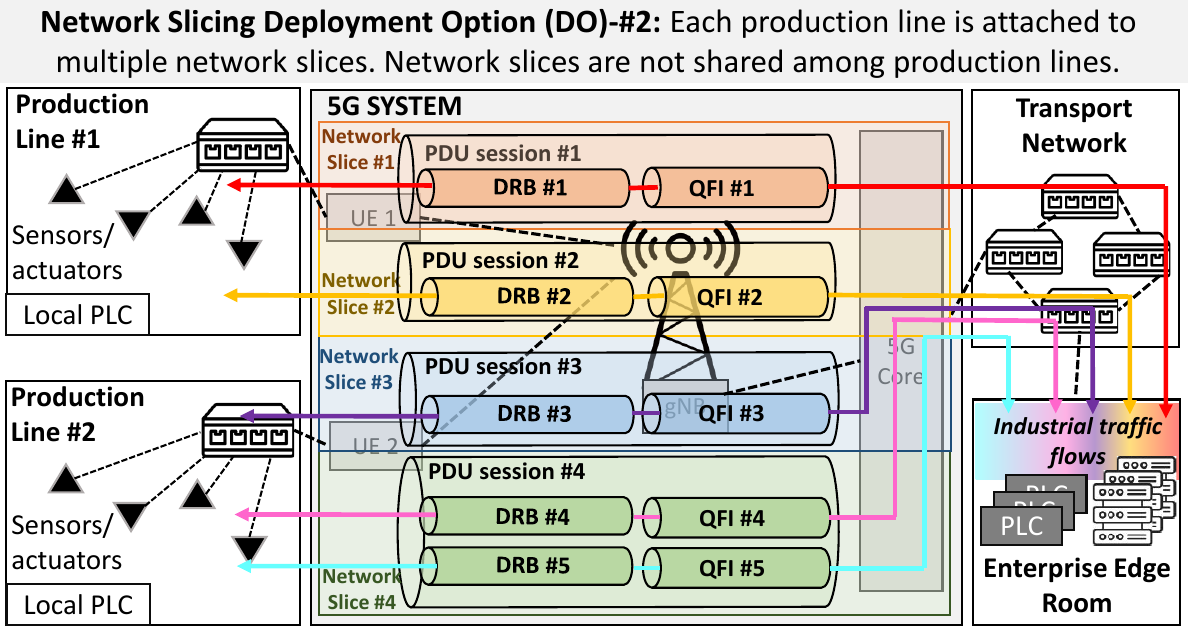}
  \end{minipage}

  \vspace{0.1em}

  \begin{minipage}[t]{\dimexpr0.5\textwidth-0.5\subfigsep\relax}
    \centering
    \includegraphics[width=0.95\linewidth]{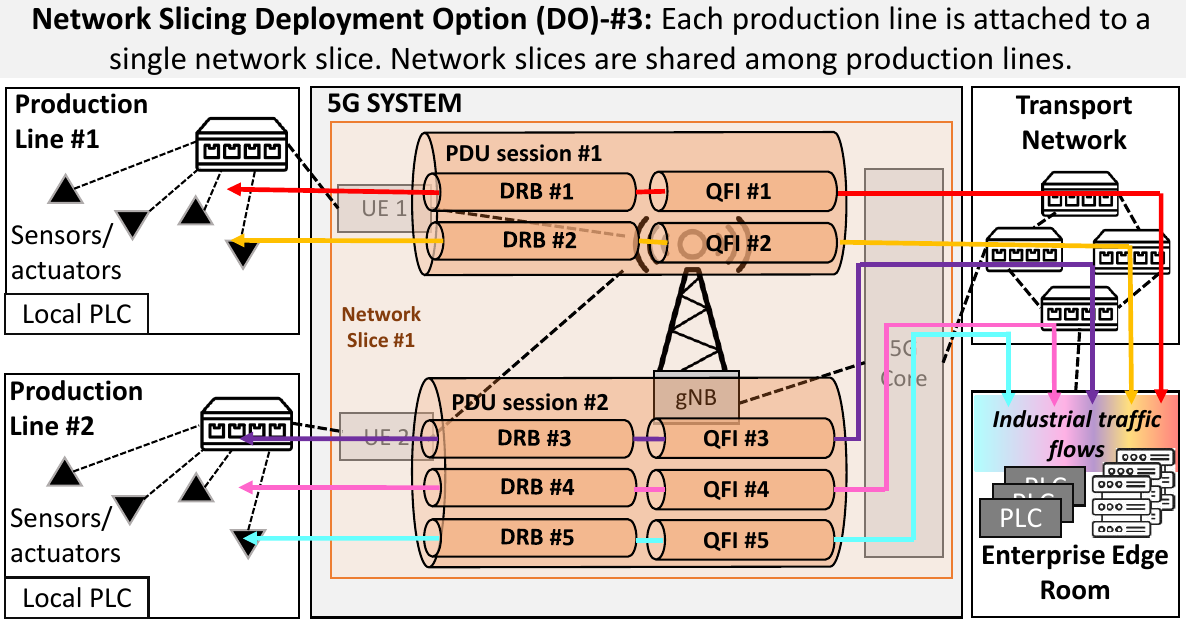}
  \end{minipage}\hspace{\subfigsep}%
  \begin{minipage}[t]{\dimexpr0.5\textwidth-0.5\subfigsep\relax}
    \centering
    \includegraphics[width=0.95\linewidth]{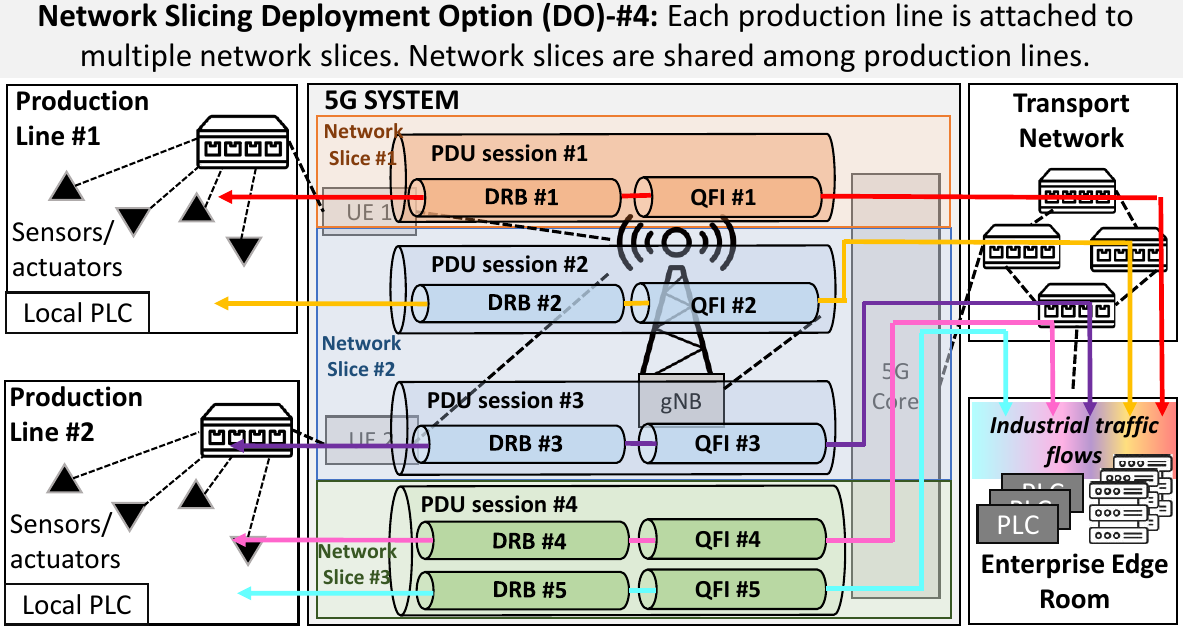}
  \end{minipage}

  \caption{Network slicing deployment options in a 5G-based industrial network. We assume production lines in a factory connect wirelessly to the \gls{5G} system, which then interfaces with the enterprise edge cloud through a transport network~\cite{5GACIA-whitepaperI}. }
  \label{fig:NetworkSlicingOptions}
\end{figure*}

We consider a single-factory industrial scenario with multiple production lines served by one \gls{5G} \gls{BS}. Each production line connects to the \gls{5G} system through a single \gls{UE} that aggregates all traffic flows of that line, consistently with gateway-based industrial integration scenarios considered by \gls{5G}-ACIA~\cite{5GACIA-whitepaperI}, where field devices exchange traffic through a 5G-connected \gls{UE}.

Industrial traffic may include control and synchronization flows, cyclic data exchanges, event-triggered messages, robotic communications, augmented-reality video, and configuration or diagnostic traffic, each with different performance requirements~\cite{Adamuz-Hinojosa2025}. Since the number and types of flows depend on the specific factory use case, we adopt a generic setting able to accommodate heterogeneous traffic mixes.

We analyze four \gls{RAN} slicing deployment options defined by two binary design dimensions: \textit{(i) whether slices are shared across production lines} and \textit{(ii) whether slicing is performed at the production-line level}. The resulting configurations are shown in Fig.~\ref{fig:NetworkSlicingOptions}. In general, heterogeneous and delay-critical flows benefit from finer-grained slicing, whereas more homogeneous or less stringent traffic can be served more efficiently through shared slices.

\textbf{Deployment Option (DO)-\#1: One Slice per Production Line, No Inter-Line Sharing}. Each production line is assigned one dedicated network slice that is not shared with other lines. Since each line is served by a single \gls{UE}, the number of slices equals the number of production lines, and each \gls{UE} establishes one \gls{PDU} session carrying all flows of that line. 
\emph{Use case}: this option provides inter-line isolation, so disturbances in one production line do not propagate to others. For example, a burst of event-triggered alarms caused by a local fault remains confined within the slice of the affected line, avoiding degradation of control accuracy in neighboring lines.

\textbf{Deployment Option (DO)-\#2: Multiple Slices per Production Line, No Inter-Line Sharing}. Each production line is attached to multiple network slices that are not shared with other lines, with each slice mapped to a different \gls{PDU} session. In the most granular case, one slice is assigned per traffic flow, although a practical alternative is to group flows by traffic category (e.g., cyclic control, video, best-effort, or diagnostics). 
\emph{Use case}: deterministic closed-loop motion control can be isolated in a dedicated slice, while visual inspection or monitoring traffic is confined to another, preventing delay inflation due to heterogeneous traffic and preserving timing determinism.

\textbf{Deployment Option (DO)-\#3: One Slice per Production Line, Shared across Production Lines}. Each production line is associated with a single network slice, which may be shared with other lines. When multiple lines share the same slice, the slice supports one \gls{PDU} session per \gls{UE}, resulting in multiple sessions anchored at the corresponding \glspl{UE}. For instance, in Fig.~\ref{fig:NetworkSlicingOptions}, Production Lines~\#1 and~\#2 share one slice comprising two \gls{PDU} sessions, one per line. 
\emph{Use case}: when several production lines generate homogeneous and non-critical traffic, slice sharing improves radio-resource efficiency and scalability. The drawback is weaker temporal isolation, which may degrade production accuracy if delay-critical traffic is aggregated.

\textbf{Deployment Option (DO)-\#4: Multiple Slices per Production Line, with Shared Slices}. Each production line is attached to multiple network slices, at least one of which is shared with other lines. Shared slices support one \gls{PDU} session per \gls{UE}, while the remaining slices remain dedicated to individual lines. 
\emph{Use case}: high-bandwidth video streams from inspection stations across multiple lines can be aggregated into a shared slice, whereas deterministic motion-control flows of robotic arms remain in dedicated slices, preserving control precision while improving overall radio-resource efficiency.

\section{System Model}\label{sec:SystemModel}
Let $\mathcal{U}$ denote the set of \glspl{UE}, where each $u \in \mathcal{U}$ represents a production line. Each \gls{UE} carries multiple traffic flows collected in $\mathcal{F}$, and each flow $f \in \mathcal{F}$ is characterized by a delay bound $W_f$ and a violation probability $\varepsilon_f$, i.e., $\mathbb{P}[w > W_f] \leq \varepsilon_f$, where $w$ denotes the radio-interface transmission delay. The subset $\mathcal{F}_u \subseteq \mathcal{F}$ contains the flows associated with \gls{UE} $u$. The \gls{BS} implements one of the slicing deployment options in Section~\ref{sec:SlicingStrategies}. The set of slices is denoted by $\mathcal{S}$, and $\mathcal{F}_s \subseteq \mathcal{F}$ denotes the flows served by slice $s \in \mathcal{S}$. Only latency- and reliability-constrained industrial traffic is considered. Since the objective is to compare \gls{RAN} slicing deployment options and the delay metric refers to the radio-interface transmission delay, our study is restricted to the wireless access segment.

\subsection{Traffic Model}
Each flow $f \in \mathcal{F}$ is modeled as a Poisson packet arrival process with average rate $\lambda_f$. Packet sizes are denoted by $d_{f,i}$, where $i \in \mathcal{N}_f^{\text{pkt}}$ indexes the set of possible sizes, and occur with probability $p_{f,i}^{\text{pkt}}$, with $\sum_{i \in \mathcal{N}_f^{\text{pkt}}} p_{f,i}^{\text{pkt}} = 1$.

Flows aggregate packets with similar delay requirements and identical source--destination pairs, potentially originating from different applications or events. At the planning time scale considered in this work, packet arrivals are modeled as a Poisson process as a tractable first-order characterization of aggregated traffic. This choice is also supported by the Palm--Khintchine theorem, according to which the superposition of multiple independent packet sources can be approximated by a Poisson process under aggregation. The model provides a common analytical basis for comparing the considered slicing deployment options, while more structured periodic or burst-driven industrial traffic patterns are left for future work.

\subsection{Radio Resource Model}
Let $N_{\text{cell}}^{\text{RB}}$ denote the total number of available \glspl{RB} in the cell. The \gls{BS} assigns $N_{s}^{\text{RB}}$ \glspl{RB} to each slice $s \in \mathcal{S}$, subject to $\sum_{s \in \mathcal{S}} N_{s}^{\text{RB}} \leq N_{\text{cell}}^{\text{RB}}$. Within slice $s$, resources are distributed among its flows, with $N_{f}^{\text{RB}}$ denoting the \gls{RB} allocation of flow $f \in \mathcal{F}_s$, such that $N_{s}^{\text{RB}} = \sum_{f \in \mathcal{F}_s} N_{f}^{\text{RB}}$. For slices serving multiple flows, a uniform Round Robin split is assumed, i.e., $N_{f}^{\text{RB}} = N_{s}^{\text{RB}} / |\mathcal{F}_s|$. Since the focus is on slice-level resource planning at long time scales, this scheduler-agnostic assumption isolates the impact of slice deployment and \gls{RB} provisioning on delay-bound satisfaction, while the design of \gls{QoS}-aware or channel-aware intra-slice schedulers is left outside the scope of this work.

\subsection{Channel Model}
The signal received by \gls{UE} $u$ on \gls{RB} $n$ is
$y_{u,n} = \sqrt{P_{u,n}^{\text{tx}} P_{u,n}^{\text{pl}}}\,\tilde{h}_{u,n} s_{u,n} + n_0$,
where $s_{u,n}$ has power $P_{u,n}^{\text{tx}}$, $P_{u,n}^{\text{pl}}$ denotes the path loss, $\tilde{h}_{u,n}=h_{u,n}e^{jv_{u,n}}$ models fast fading, and $n_0$ is AWGN. The instantaneous \gls{SNR} at \gls{UE} $u$ on \gls{RB} $n$ is $\gamma_{u,n}=\bar{\gamma}_{u,n}|h_{u,n}|^2$, with average $\bar{\gamma}_{u,n}=P_{u,n}^{\text{tx}} P_{u,n}^{\text{pl}}/N_0$, where $N_0$ is the noise power. Since production lines are deployed at fixed locations within the factory and the focus is on slice planning over long time scales, $\bar{\gamma}_{u,n}$ is assumed time-invariant within each planning window. Moreover, the industrial cell is assumed to operate under controlled spectrum conditions, so inter-cell interference in downlink is neglected.

As the \gls{BS} serves the factory from a single location and line-of-sight cannot be guaranteed due to machinery and metallic obstructions, fast fading is modeled as Rayleigh. Under this assumption, the instantaneous \gls{SNR} follows an exponential distribution with \gls{PDF}
$f_{\gamma_u}(\gamma)=\frac{1}{\bar{\gamma}_u}\exp(-\gamma/\bar{\gamma}_u)$, which no longer depends on the \gls{RB} index. Accordingly, the probability that \gls{UE} $u$ employs \gls{MCS} $m\in\mathcal{M}$ is
$p_{u,m}=\int_{\gamma_m^{\min}}^{\gamma_m^{\max}}\frac{1}{\bar{\gamma}_u}\exp(-\gamma/\bar{\gamma}_u)\,d\gamma$,
where $\mathcal{M}$ denotes the set of available \gls{MCS} levels, each selected over its corresponding \gls{SNR} range $[\gamma_m^{\min},\gamma_m^{\max})$.

\section{Network Slice Planner}

\subsection{Packet Delay Modeling via Stochastic Network Calculus}\label{sec:EstimationSNC}
We use \gls{SNC} to derive per-flow packet delay bounds. Let $A_f(\tau,t)$ and $S_f(\tau,t)$ denote the cumulative arrival and service processes over $(\tau,t]$ for flow $f \in \mathcal{F}$. In \gls{SNC}, these processes are represented through probabilistic envelopes,
$\mathbb{P}[A_f(\tau,t)>\alpha_f(\tau,t)]\le \varepsilon_{A_f}$ and
$\mathbb{P}[S_f(\tau,t)<\beta_f(\tau,t)]\le \varepsilon_{S_f}$,
where $\alpha_f(\tau,t)$ and $\beta_f(\tau,t)$ are the arrival and service envelopes, and $\varepsilon_{A_f}$ and $\varepsilon_{S_f}$ denote overflow and deficit probabilities. These envelopes are derived from the statistical characterization of the underlying arrival and service processes and capture traffic variability and wireless-service fluctuations, including packet-size randomness, traffic intensity, and variations in the effective \gls{RAN} service rate. This probabilistic framework enables delay provisioning under a target violation probability. The delay bound $W_f$ is given by the maximum horizontal deviation between $\alpha_f(\tau,t)$ and $\beta_f(\tau,t)$, assuming that the long-term service rate exceeds the arrival rate. Further details on \gls{SNC} fundamentals can be found in~\cite{Fidler2010,Fidler1}.

\textbf{\gls{SNC} Methodology.} The estimation of $W_f$ using \gls{SNC} follows three main steps~\cite{Fidler2010,Fidler1}.

\textit{1) \glspl{MGF}:} The arrival and service processes $A_f(\tau,t)$ and $S_f(\tau,t)$ are first characterized through their \glspl{MGF}. For a random variable $X$, the \gls{MGF} is $M_X(\theta)=\mathbb{E}[e^{\theta X}]$, where $\theta>0$ is a tunable parameter.

\textit{2) Exponential bounds:} By applying Chernoff’s inequality, the cumulative arrival and service processes are upper- and lower-bounded by affine envelopes $\alpha_f(\tau,t)$ and $\beta_f(\tau,t)$, parameterized by $(\rho_{A_f},\sigma_{A_f})$ and $(\rho_{S_f},\sigma_{S_f})$. These parameters are obtained by fitting the corresponding \glspl{MGF} to exponential bounds.

\textit{3) Delay bound derivation:} Once the envelopes are defined, the delay bound $W_f$ is given by the maximum horizontal deviation between $\alpha_f$ and $\beta_f$. In practice,
$W_f = (b_{A_f}+b_{S_f})/(\rho_{S_f}-\delta)$, where
$b_{A_f}=\sigma_{A_f}-\left[\ln(\varepsilon_{A_f})+\ln(1-\exp(-\theta\delta))\right]/\theta$ and
$b_{S_f}=\sigma_{S_f}-\left[\ln(\varepsilon_{S_f})+\ln(1-\exp(-\theta\delta))\right]/\theta$
are burst terms, and $\delta>0$ is a tunable parameter. The stability condition $\rho_{S_f}-\delta>\rho_{A_f}+\delta$ must hold, and $(\theta,\delta)$ are optimized to minimize $W_f$ under the target violation probability $\varepsilon_f=\varepsilon_{A_f}+\varepsilon_{S_f}$.

Next, we summarize the computation of the arrival and service processes and the resulting delay bound. These expressions are minor adaptations of validated \gls{SNC} formulations~\cite{Adamuz2022,Adamuz2024} and provide a common basis for comparing \gls{RAN} slicing deployment options.

\textbf{Arrival Process for a Traffic Flow.} Packet arrivals follow a Poisson process with rate $\lambda_f$, while packet sizes take discrete values $\{d_{f,i}\}$ with \gls{PMF} $p^{\text{pkt}}_{f,i}$. Over $(\tau,t]$, the accumulated arrivals form a compound Poisson process with \gls{MGF}
\begin{equation}
M_{A_f}(\theta)=\exp\!\Big(\lambda_f (t-\tau)\,[M_{L_f}(\theta)-1]\Big),
\end{equation}
where $M_{L_f}(\theta)=\sum_{i\in \mathcal{N}_f^{\text{pkt}}} p^{\text{pkt}}_{f,i}\exp(\theta d_{f,i})$. This expression captures the variability induced by both random arrivals and packet-size fluctuations. The resulting affine arrival envelope $\alpha_f(\tau,t)$ is characterized by
\begin{equation}
\rho_{A_f}(\theta)=\lambda_f\,[M_{L_f}(\theta)-1]/\theta, \quad \sigma_{A_f}(\theta)=0 .
\end{equation}

\textbf{Service Process for a Traffic Flow.} In each \gls{TTI}, the \gls{BS} allocates $N_f^{\text{RB}}$ \glspl{RB} to flow $f \in \mathcal{F}$. The number of bits served per \gls{RB} is $N_{\text{SC}}\Delta_f t_{\text{slot}}\eta$, where $N_{\text{SC}}$ is the number of subcarriers per \gls{RB}, $\Delta_f$ is the subcarrier spacing, $t_{\text{slot}}$ is the slot duration, and $\eta$ is the spectral efficiency selected by \gls{MCS} adaptation. Let $p_{f,m}=\mathbb{P}\{\eta=\eta_m\}$ denote the probability of using \gls{MCS} $m\in\mathcal{M}$, induced by the per-\gls{RB} \gls{SNR} distribution; since each flow $f$ is associated with a specific \gls{UE} $u$, $p_{f,m}=p_{u,m}$ if $f$ belongs to $u$. The corresponding negative \gls{MGF} of the per-\gls{RB} service is
$M_{\eta}(-\theta)=\sum_{m\in\mathcal{M}} p_{f,m}\exp\!\left(-\theta N_{\text{SC}}\Delta_f t_{\text{slot}}\eta_m\right)$,
where evaluation at $-\theta$ is standard in the exponential domain of the service envelope. Assuming independence across \glspl{RB} and \glspl{TTI}, the cumulative service process $S_f(\tau,t]$ has negative \gls{MGF}
\begin{equation}
M_{S_f}(-\theta)=\exp\!\left(\ln\!\left[M_{C_f}(-\theta)\right]\frac{t-\tau}{t_{\text{slot}}}\right),
\label{eq:MGF_service}
\end{equation}
with
$M_{C_f}(-\theta)=\left(\sum_{m\in\mathcal{M}} p_{f,m}\exp\!\left(-\theta N_{\text{SC}}\Delta_f t_{\text{slot}}\eta_m\right)\right)^{N_f^{\text{RB}}}$,
yielding $\rho_{S_f}(\theta)= -\ln\!\left(M_{C_f}(-\theta)\right)/(\theta t_{\text{slot}})$ and $\sigma_{S_f}(\theta)=0$. This formulation captures the variability of the wireless service across \glspl{TTI} and allocated \glspl{RB} due to random \gls{MCS} selection induced by channel fading.

\textbf{Delay Bound for a Traffic Flow.} The violation budget is evenly split between arrival and service envelopes, i.e., $\varepsilon_{A_f}=\varepsilon_{S_f}=\varepsilon_f/2$. Under this split and the envelope construction above, the per-flow delay bound is
$W_f(\theta,\delta)=
-\frac{2\left[\ln(\varepsilon_f/2)+\ln\!\left(1-e^{-\theta\delta}\right)\right]}
{\theta\left(\rho_{S_f}(\theta)-\delta\right)}$.
The parameters $(\theta,\delta)$ are obtained by solving
\begin{equation}
\min_{\theta>0,\ \delta>0}\ \ W_f(\theta,\delta)
\quad \text{s.t.}\quad 
\rho_{S_f}(\theta)-\delta>\rho_{A_f}(\theta)+\delta,
\label{eq:snc_opt}
\end{equation}
which enforces the stability condition required for a finite delay bound.

Problem~\eqref{eq:snc_opt} is non-convex in both the objective function and the feasible region, leading to multiple local minima and making exhaustive search impractical. To address this, we adapt the heuristic proposed in~\cite{Adamuz2024}.

\subsection{Problem Formulation}
For each flow $f \in \mathcal{F}$, the delay bound $W_f$ is estimated using \gls{SNC} (Section~\ref{sec:EstimationSNC}) and compared against its target value $W_f^{\mathrm{obj}}$. We define the normalized delay bound as $W_f^{\mathrm{norm}} = W_f / W_f^{\mathrm{obj}}$, and a flow satisfies its \gls{QoS} requirements if $W_f^{\mathrm{norm}} \le 1$.

The \gls{RB} allocation problem is formulated in Eq.~\eqref{eq:obj_function}. The optimization variables are the slice-level \gls{RB} allocations $\{N_s^{RB}\}_{s \in \mathcal{S}}$, which indirectly determine the per-flow delay bounds. An auxiliary variable $\zeta$ captures the maximum normalized delay bound across all flows. The goal is to characterize feasible slice-level allocations under per-flow \gls{QoS} and cell-capacity constraints, while favoring margin-tight solutions in which the worst-case normalized delay remains as close as possible to unity without exceeding it:
\begin{equation}
\begin{aligned}
    \max_{\{N_s^{RB}\},\, \zeta} \quad & 
    \zeta = \max_{f \in \mathcal{F}} W_f^{\mathrm{norm}} \\
    \text{s.t.} \quad 
    & \zeta \le 1,\ \ \sum_{s\in\mathcal{S}} N_s^{RB} \le N_{cell}^{RB}.
\end{aligned}
\label{eq:obj_function}
\end{equation}
The first constraint enforces compliance with the target delay bounds, while the second limits the total allocated \glspl{RB} to the available cell capacity. In this formulation, maximizing $\zeta$ under $\zeta \le 1$ favors margin-tight slice-level allocations, where radio resources are provisioned just sufficiently to satisfy all per-flow delay requirements without over-provisioning.

\subsection{Heuristic Solution}\label{sec:heuristic}
The optimization problem is non-convex and NP-hard, which precludes the use of exact solution methods. We therefore propose a heuristic iterative algorithm structured into two phases. For a given amount of radio resources, Phase~A first reduces the worst-case normalized delay bound through inter-slice \gls{RB} reallocation. Then, starting from a feasible allocation, Phase~B progressively removes \glspl{RB} while preserving $\zeta \le 1$, thereby obtaining a margin-tight solution without over-provisioning.

\textbf{Phase A. Delay Balancing.}
The algorithm starts from an equal \gls{RB} allocation among slices. At each iteration, the flow with the highest normalized delay bound $W_f^{\mathrm{norm}}$ and the flow with the lowest $W_f^{\mathrm{norm}}$ belonging to a different slice are identified. These slices are denoted as the worst and best slices, respectively. One \gls{RB} is reallocated from the best slice to the worst slice, provided that the donor slice has more than one \gls{RB}. After each reallocation, the per-flow delay bounds $W_f$, $\forall f \in \mathcal{F}$, are re-estimated using the \gls{SNC} model. The reallocation is accepted only if it reduces the objective value $\zeta$ in Eq.~\eqref{eq:obj_function}; otherwise, Phase~A terminates. Since each iteration performs a single unitary \gls{RB} reallocation and triggers a full recomputation of per-flow delay bounds, and the number of admissible reallocations is upper-bounded by the total number of available \glspl{RB}, $N_{\mathrm{cell}}^{\mathrm{RB}}$, the computational complexity of Phase~A is $\mathcal{O}(N_{\mathrm{cell}}^{\mathrm{RB}} |\mathcal{F}|)$. At convergence, Phase~A yields the minimum attainable worst-case normalized delay bound for the given amount of radio resources. If the resulting allocation satisfies $W_f^{\mathrm{norm}} \le 1$ for all flows, the algorithm proceeds to Phase~B; otherwise, it terminates, as the available spectrum is insufficient to meet all delay targets.

\textbf{Phase B. Resource Tightening.}
Phase~B reduces the total number of allocated \glspl{RB} while maintaining per-flow delay feasibility. At each iteration, one \gls{RB} is tentatively removed from each slice $s$ with $N_s^{\mathrm{RB}}>1$, and the resulting allocation is evaluated by recomputing all per-flow delay bounds using the \gls{SNC} model, assuming uniform intra-slice sharing. Among the feasible candidates, i.e., those for which the maximum normalized delay satisfies $\zeta \le 1$, the slice whose \gls{RB} removal yields the largest $\zeta$ is selected and the removal is applied. This process is repeated until no further \gls{RB} can be removed without violating delay constraints. Each iteration evaluates up to $|\mathcal{S}|$ candidates, and at most $N_{\mathrm{cell}}^{\mathrm{RB}}$ successful removals are possible, resulting in a computational complexity of $\mathcal{O}(N_{\mathrm{cell}}^{\mathrm{RB}} |\mathcal{S}| |\mathcal{F}|)$.

\textbf{Computational Complexity.}
Based on the above analysis, the worst-case computational complexity of the proposed heuristic is $\mathcal{O}(N_{\mathrm{cell}}^{\mathrm{RB}} |\mathcal{S}| |\mathcal{F}|)$. In practice, the execution time is often lower, as either Phase~A or Phase~B may terminate early depending on the slicing configuration and the available radio resources.

\subsection{Window-Based Operation of the SNC-Based Slice Planner}
\label{subsec:window_based_planner}

The proposed slice planner operates over a finite \emph{planning window}, during which the statistical descriptors of traffic are assumed to be stationary. The overall operation time is partitioned into a sequence of planning windows $\mathcal{W}_k = [t_k,\, t_k + T_{\mathrm{w}})$, $k \geq 0$, where $T_{\mathrm{w}}$ denotes the window duration. Within each window $\mathcal{W}_k$, the planner computes a slice-level \gls{RB} allocation based on a fixed set of traffic statistics.

For each traffic flow $f \in \mathcal{F}$, the system behavior within window $\mathcal{W}_k$ is characterized by the tuple
$(\lambda_f^{(k)}, \{p_{f,i}^{\mathrm{pkt},(k)}\}_{i \in \mathcal{N}_f^{\mathrm{pkt}}}, \{p_{u,m}\}_{m \in \mathcal{M}})$,
which determines the arrival and service envelopes used by the \gls{SNC}-based model to compute the per-flow delay bound $W_f^{(k)}$. The stationarity assumption applies within a single planning window, while traffic statistics are allowed to evolve across windows.

In the considered industrial \gls{RAN} scenario, \glspl{UE} are static and associated with fixed production lines. As a result, the average \gls{SNR} experienced by each \gls{UE}, $\bar{\gamma}_u$, is assumed to remain constant within each planning window. Consequently, the \gls{MCS} selection \gls{PMF} $\{p_{u,m}\}$ is also assumed to remain invariant within each planning window. In contrast, traffic characteristics such as average arrival rates or packet size distributions may evolve over time scales of minutes. These properties motivate a slice planning operation at a non-\gls{RT} time scale.

Accordingly, the proposed slice planner operates at the Non-\gls{RT} \gls{RIC} within the \gls{O-RAN} architecture and may be implemented as an \emph{rApp}~\cite{survey-O-RAN}. The planner computes slice-level \gls{RB} budgets from long-term traffic statistics, and these budgets are enforced by \gls{DU} MAC schedulers operating at the \gls{TTI} scale. A window-based \gls{SNC}-based control approach was previously introduced in~\cite{Adamuz2024} for near-\gls{RT} per-user resource allocation in public broadband networks without slicing. In contrast, the planner proposed in this paper adopts the \gls{SNC} model developed herein, operates at the Non-\gls{RT} \gls{RIC}, and extends that framework to slice-level resource planning tailored to industrial \gls{RAN} scenarios.

Under this window-based, Non-\gls{RT} operation, the slice planning workflow proceeds as follows. Long-term traffic statistics are aggregated at the Non-\gls{RT} \gls{RIC} via the O1 interface, yielding estimates of $\lambda_f^{(k)}$ and $\{p_{f,i}^{\mathrm{pkt},(k)}\}$ over $\mathcal{W}_k$. These statistics are used by an \gls{SNC}-based slice planning \emph{rApp} to compute per-flow delay bounds and solve Eq.~\eqref{eq:obj_function}, obtaining the slice-level \gls{RB} allocation $\{N_s^{\mathrm{RB},(k)}\}_{s \in \mathcal{S}}$. The resulting allocation is translated into slice-level policies and conveyed to the near-\gls{RT} \gls{RIC} via the A1 interface. These policies are subsequently enforced through near-\gls{RT} control actions delivered over the E2 interface to the \gls{gNB}, where they are applied by \gls{DU} MAC schedulers operating at the \gls{TTI} scale.

By iterating this procedure over successive planning windows, the slice planner periodically updates slice-level resource allocations based on slowly varying traffic statistics, enabling long-term adaptation to changing operating conditions while preserving the \gls{SNC}-based delay guarantees within each planning interval.

\section{Performance Results}\label{sec:perf_results}
\subsection{Experimental Setup}

We consider a single-cell industrial scenario with one centrally located \gls{BS} serving three production lines placed at increasing distances, each demanding three downlink flows. The evaluation is simulation-based and relies on the proposed \gls{SNC}-based analytical model and its Python implementation. The \gls{BS} operates using 5G~NR with subcarrier spacing $\Delta f=60$\,kHz and carrier frequency $f_c=4.7$\,GHz. Two bandwidth configurations are considered, corresponding to $N_{\mathrm{cell}}^{\mathrm{RB}}\in\{65,135\}$. The transmit power is $P_{\mathrm{tx}}=24$\,dBm and the thermal noise density is $N_0=-174$\,dBm/Hz. Packets have fixed size $L=512$\,bits. Flow arrival rates, delay requirements, and distances to the \gls{BS} are summarized in Table~\ref{tab:ExpSetup}.

We evaluate five deployment options. In the baseline (DO-\#0), no slicing is applied and all flows are multiplexed within a single slice, $\mathcal{S}_1=\{f_1,\dots,f_9\}$. In DO-\#1, one slice is assigned per production line, with $\mathcal{S}_1=\{f_1,f_2,f_3\}$, $\mathcal{S}_2=\{f_4,f_5,f_6\}$, and $\mathcal{S}_3=\{f_7,f_8,f_9\}$. In DO-\#2, one dedicated slice is assigned per flow, i.e., $\mathcal{S}_f=\{f\}$ for all $f \in \mathcal{F}$. In DO-\#3, flows from the first and second production lines, which have less restrictive delay-bound requirements, are grouped into a shared slice $\mathcal{S}_1=\{f_1,\dots,f_6\}$, while flows from the third line, characterized by more stringent delay-bound requirements, are assigned to a dedicated slice $\mathcal{S}_2=\{f_7,f_8,f_9\}$. Finally, DO-\#4 groups flows with $W_f^{\mathrm{obj}}=1$\,ms into $\mathcal{S}_1=\{f_3,f_6\}$, flows with $W_f^{\mathrm{obj}}=0.5$\,ms into $\mathcal{S}_2=\{f_1,f_2,f_4,f_5\}$, and assigns the most delay-critical flows to dedicated slices $\mathcal{S}_3=\{f_7\}$, $\mathcal{S}_4=\{f_8\}$, and $\mathcal{S}_5=\{f_9\}$.

All experiments were executed on a machine equipped with 16\,GB of RAM and a quad-core Intel Core i7-7700HQ processor running at 2.80\,GHz.

\begin{table}[t!]
\centering
\caption{Per-UE traffic and QoS requirements.}
\label{tab:ExpSetup}
\scriptsize
\begin{tabular}{ccccc}
\toprule
UE &
Flows &
$\lambda_f$ (pkt/s) &
$W_f^{obj}$ (ms) &
\makecell{Distance\\to BS (m)} \\
\midrule
UE~\#1 &
$f_1,f_2,f_3$ &
2000, 3000, 1500 &
0.5, 0.5, 1.0 &
80 \\
UE~\#2 &
$f_4,f_5,f_6$ &
5000, 6600, 4500 &
0.5, 0.5, 1.0 &
200 \\
UE~\#3 &
$f_7,f_8,f_9$ &
9000, 11000, 8000 &
0.2, 0.2, 0.5 &
350 \\
\bottomrule
\end{tabular}
\end{table}

\subsection{Comparison of Network Slicing Strategies}
\begin{figure}[t!]
    \centering
    \includegraphics[width=0.88\columnwidth]{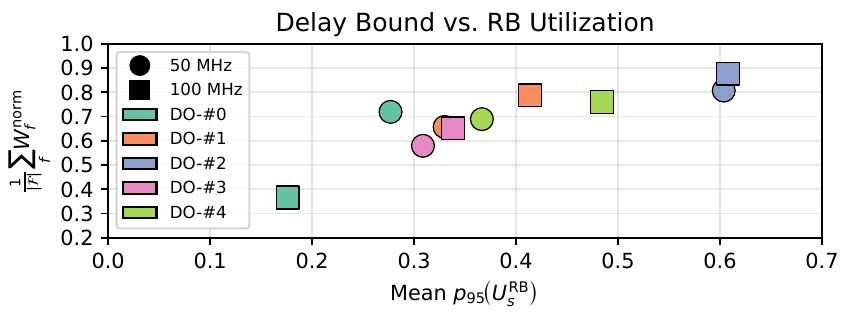}
    \caption{Slice-level RB utilization and delay--utilization trade-off across deployment options.}
    \label{fig:efficiency}
\end{figure}

This section compares the considered \gls{RAN} slicing deployment options under a common slice-planning framework, focusing on per-flow delay guarantees and radio resource utilization. Table~\ref{tab:comp_scenarios} reports the normalized delay bounds for all flows, Table~\ref{tab:FinalRBsslices} summarizes the slice-level \gls{RB} allocations obtained after convergence of the proposed planner, and Fig.~\ref{fig:efficiency} illustrates the trade-off between \gls{RB} utilization and delay performance.

\textbf{Delay satisfaction.}  
The impact of the slicing strategy is best understood by distinguishing between abundant and scarce bandwidth regimes. Under the 100\,MHz configuration ($N_{cell}^{RB}=135$), all deployment options satisfy the per-flow delay requirements, including the baseline without slicing (DO-\#0). In this regime, flows $f_7$ and $f_8$, which have the highest arrival rates and most stringent delay requirements, consistently exhibit the largest normalized delay bounds, while still remaining below unity.

Under radio resource scarcity ($N_{cell}^{RB}=65$), the effect of the slicing criterion becomes evident. The baseline configuration (DO-\#0) exhibits severe delay violations for flows $f_7$ and $f_8$. Similar violations appear in DO-\#1, where isolation is enforced at the production-line level but flows with heterogeneous delay requirements still share the same slice. In particular, the violations originate in the slice associated with UE~\#3, which aggregates two highly delay-critical flows together with a third flow subject to a more relaxed delay bound, thus providing insufficient protection for the most stringent traffic. DO-\#2 provides the most robust delay performance under scarce bandwidth. By assigning one dedicated slice per flow, the planner can closely match allocated \glspl{RB} to individual delay requirements, yielding the lowest worst-case normalized delay bounds and avoiding violations across all flows. Slice-sharing configurations exhibit a differentiated behavior. In DO-\#3, delay violations persist for flows $f_7$ and $f_8$, but the corresponding normalized delay bounds remain close to unity, indicating that grouping flows according to relative delay criticality partially alleviates the impact of resource scarcity. In DO-\#4, violations also occur under limited bandwidth, but they no longer affect the most delay-critical flows; instead, they shift towards flows with intermediate delay requirements, reflecting the prioritization of stringent traffic within the slice-level resource allocation.

\begin{table}[t!]
\centering
\caption{Normalized delays $W_f^{norm}$ for all flows.}
\resizebox{1.0\columnwidth}{!}{
\begin{tabular}{lccccccccc}
\toprule
Scenario ($N_{cell}^{RB}$) & $f_1$ & $f_2$ & $f_3$ & $f_4$ & $f_5$ & $f_6$ & $f_7$ & $f_8$ & $f_9$ \\
\midrule
DO-\#0 (65)  & 0.26 & 0.25 & 0.14 & 0.47 & 0.50 & 0.22 & \textbf{1.88} & \textbf{2.02} & 0.73 \\
DO-\#0 (135) & 0.16 & 0.15 & 0.08 & 0.23 & 0.25 & 0.12 & 0.91 & 0.99 & 0.40 \\
\midrule
DO-\#1 (65)  & 0.29 & 0.30 & 0.17 & 0.39 & 0.50 & 0.22 & \textbf{1.59} & \textbf{1.72} & 0.73 \\
DO-\#1 (135) & 0.84 & 0.93 & 0.66 & 0.87 & 0.92 & 0.56 & 0.91 & 0.99 & 0.40 \\
\midrule
DO-\#2 (65)  & 0.77 & 0.80 & 0.57 & 0.92 & 0.78 & 0.67 & 0.90 & 0.98 & 0.87 \\
DO-\#2 (135) & 0.84 & 0.93 & 0.66 & 0.87 & 0.92 & 0.84 & 0.91 & 0.99 & 0.93 \\
\midrule
DO-\#3 (65)  & 0.35 & 0.36 & 0.17 & 0.69 & 0.78 & 0.35 & \textbf{1.03} & \textbf{1.05} & 0.43 \\
DO-\#3 (135) & 0.46 & 0.52 & 0.23 & 0.87 & 0.92 & 0.56 & 0.91 & 0.99 & 0.40 \\
\midrule
DO-\#4 (65)  & 0.54 & 0.80 & 0.17 & \textbf{1.39} & \textbf{1.05} & 0.27 & 0.80 & 0.87 & 0.31 \\
DO-\#4 (135) & 0.46 & 0.52 & 0.39 & 0.87 & 0.92 & 0.84 & 0.91 & 0.99 & 0.93 \\
\bottomrule
\end{tabular}
}
\label{tab:comp_scenarios}
\end{table}

\begin{table}[t!]
\centering
\caption{Allocated RBs per Slice after Algorithm Execution.}
\scriptsize
\setlength{\tabcolsep}{3pt}
\resizebox{1.0\columnwidth}{!}{
\begin{tabular}{@{}lrr lrr@{}}
\toprule
\multicolumn{3}{c}{\textbf{DO-\#1}} & \multicolumn{3}{c}{\textbf{DO-\#3}} \\
\cmidrule(lr){1-3}\cmidrule(lr){4-6}
Slice & 50\,MHz & 100\,MHz & Slice & 50\,MHz & 100\,MHz \\
\midrule
S1 $\{f_1,f_2,f_3\}$ & 20 &  8  & S1 $\{f_1,\dots,f_6\}$ & 33 & 29 \\
S2 $\{f_4,f_5,f_6\}$ & 22 & 14  & S2 $\{f_7,f_8,f_9\}$   & 32 & 41 \\
S3 $\{f_7,f_8,f_9\}$ & 23 & 41  &                        &    &    \\
\midrule
\textbf{Total Allocated RBs} & \textbf{65} & \textbf{63} &
\textbf{Total Allocated RBs} & \textbf{65} & \textbf{70} \\
\midrule
\multicolumn{3}{c}{\textbf{DO-\#2}} & \multicolumn{3}{c}{\textbf{DO-\#4}} \\
\cmidrule(lr){1-3}\cmidrule(lr){4-6}
Slice & 50\,MHz & 100\,MHz & Slice & 50\,MHz & 100\,MHz \\
\midrule
S1 $\{f_1\}$ &  3 &  3 & S1 $\{f_3,f_6\}$          & 12 &  6 \\
S2 $\{f_2\}$ &  3 &  3 & S2 $\{f_1,f_2,f_4,f_5\}$  & 14 & 20 \\
S3 $\{f_3\}$ &  2 &  2 & S3 $\{f_7\}$              & 13 & 14 \\
S4 $\{f_4\}$ &  4 &  5 & S4 $\{f_8\}$              & 13 & 14 \\
S5 $\{f_5\}$ &  5 &  5 & S5 $\{f_9\}$              & 13 &  7 \\
S6 $\{f_6\}$ &  3 &  3 &                            &    &    \\
S7 $\{f_7\}$ & 12 & 14 &                            &    &    \\
S8 $\{f_8\}$ & 12 & 14 &                            &    &    \\
S9 $\{f_9\}$ &  6 &  7 &                            &    &    \\
\midrule
\textbf{Total Allocated RBs} & \textbf{50} & \textbf{56} &
\textbf{Total Allocated RBs} & \textbf{65} & \textbf{61} \\
\bottomrule
\end{tabular}
}
\label{tab:FinalRBsslices}
\end{table}

\textbf{Slice-level RB allocation.}  
The slice-level \gls{RB} allocations in Table~\ref{tab:FinalRBsslices} further clarify how the different strategies react to resource availability. Under spectrum scarcity ($N_{cell}^{RB}=65$), all deployment options except DO-\#2 allocate the full available \glspl{RB}, indicating that these configurations require exhausting the entire spectrum budget to satisfy delay constraints. In contrast, DO-\#2 reduces the total allocation to 50~RBs, showing that per-flow slicing enables a tighter match between allocated radio resources and individual delay-bound requirements.

When bandwidth increases to $N_{cell}^{RB}=135$, the planner reduces the total number of allocated \glspl{RB} in all deployment options, reflecting the additional flexibility introduced by a larger spectrum budget. DO-\#2 consistently achieves the lowest total \gls{RB} allocation, confirming that per-flow slicing allows a finer tuning of slice budgets and avoids over-provisioning. The remaining deployment options allocate a larger fraction of the available resources because their slice definitions aggregate multiple flows and therefore require higher slice-level budgets to accommodate heterogeneous traffic demands.

\textbf{Resource utilization vs delay bound.}  
Fig.~\ref{fig:efficiency} illustrates the relationship between slice-level \gls{RB} utilization and delay bounds. For each deployment option, it reports the average over slices of the 95th percentile of the utilization distribution, i.e., $\frac{1}{|\mathcal{S}|}\sum_{s\in\mathcal{S}} p_{95}\!\left\{U^{\mathrm{RB}}_{s}(t)\right\}$, against the average normalized delay bound across flows, $\frac{1}{|\mathcal{F}|}\sum_{f\in\mathcal{F}} W^{\mathrm{norm}}_{f}$.

Desirable operating points are those where slices make effective use of the assigned \glspl{RB} while operating close to the target delay bound. Under this criterion, the deployment options exhibit clearly differentiated behaviors. DO-\#2 consistently achieves operating points closest to this regime under both bandwidth configurations, as per-flow slicing allows the planner to finely match allocated \glspl{RB} to individual delay requirements. DO-\#4 follows, since its delay-aware slicing structure isolates the most delay-critical flows while allowing limited aggregation among flows with comparable delay requirements, thus preserving a good balance between utilization and delay tightness. In DO-\#3, grouping multiple flows with heterogeneous traffic and delay characteristics within the same slice reduces the planner’s ability to simultaneously achieve high utilization and tight delay operation. This effect is more pronounced in DO-\#1, where all flows within a production line share a slice regardless of their delay requirements. Finally, DO-\#0 exhibits the least favorable behavior, as multiplexing all flows into a single slice prevents effective differentiation.

\subsection{Execution Time Analysis}
\label{subsec:execution_time}

We evaluate the execution time of the proposed \emph{network slice planner} for the two considered bandwidth configurations, $N_{\mathrm{cell}}^{\mathrm{RB}}\in\{65,135\}$. Table~\ref{tab:execution_time} summarizes the convergence behavior and execution time.

\begin{table}[t!]
\centering
\caption{Execution time of the network slice planner.}
\label{tab:execution_time}
\scriptsize
\resizebox{\columnwidth}{!}{%
\begin{tabular}{cccccc}
\toprule
\makecell{Deployment\\Option} &
$N_{\text{cell}}^{\text{RB}}$ &
Phase A &
Phase B &
\makecell{Total\\Iterations} &
\makecell{Execution time\\(ms)} \\
\midrule
DO-\#0 & 65 (135) & 1 (1)  & 0 (10) & 1 (11)  & 149.1 (919.4) \\
DO-\#1 & 65 (135) & 3 (2)  & 0 (72) & 3 (74)  & 487.2 (15949.9) \\
DO-\#2 & 65 (135) & 11 (2) & 15 (79)& 26 (81) & 10821.9 (54587.8) \\
DO-\#3 & 65 (135) & 1 (2)  & 0 (65) & 1 (67)  & 304.1 (11147.7) \\
DO-\#4 & 65 (135) & 2 (2)  & 0 (74) & 2 (76)  & 595.1 (28851.0) \\
\bottomrule
\end{tabular}%
}
\end{table}

\textbf{Low-bandwidth regime ($N_{\mathrm{cell}}^{\mathrm{RB}}=65$).} Most deployment options converge in well below $1$\,s, with execution times between approximately $150$ and $600$\,ms for DO-\#0, DO-\#1, DO-\#3, and DO-\#4. In these cases, the limited RB budget significantly constrains the number of feasible slice-level \gls{RB} reallocations, reducing the search space explored by the planner and enabling convergence after only a few iterations. The main exception is DO-\#2, where per-flow slicing induces more iterations and leads to an execution time on the order of $10$\,s.

\textbf{High-bandwidth regime ($N_{\mathrm{cell}}^{\mathrm{RB}}=135$).} All deployment options exhibit a clear increase in both iterations and execution time. In DO-\#0, the increase remains moderate, with convergence reached in $11$ iterations and just under $1$\,s. Here, Phase~A is trivial and the execution time is dominated by Phase~B, which progressively removes \glspl{RB} from the single slice while preserving equal per-flow allocation until the minimum \gls{RB} budget satisfying all delay bounds is obtained. DO-\#1 and DO-\#3 require between $65$ and $74$ iterations and execution times on the order of $10$--$16$\,s, as slice sharing across multiple flows and production lines leads to many Phase~B tightening steps. Finally, fine-grained slicing strategies such as DO-\#2 and DO-\#4 yield the highest execution times, reaching several tens of seconds, since the larger RB budget combined with per-flow or near per-flow slice definitions substantially increases the number of feasibility-preserving RB removals in Phase~B.

\subsection{Scalability Analysis}
\label{subsec:scalability}

\begin{figure}[t!]
    \centering
    \includegraphics[width=0.9\columnwidth]{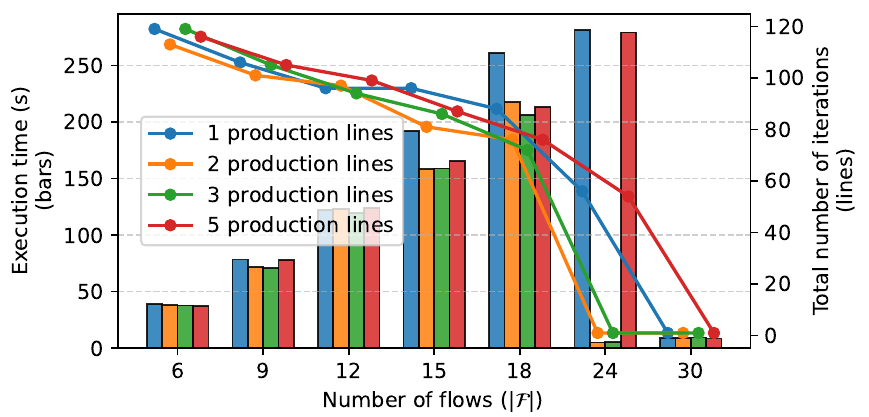}
    \caption{Execution time of the \textit{network slice planner} as a function of the number of flows $|\mathcal{F}|$ and production lines.}
    \label{fig:scalability}
\end{figure}

We evaluate the scalability of the proposed solution under $N_{\mathrm{cell}}^{\mathrm{RB}} = 135$~\glspl{RB}, focusing on DO-\#2, which is the most computationally demanding configuration. Fig.~\ref{fig:scalability} reports the execution time as a function of the total number of flows $|\mathcal{F}|$ for different numbers of production lines. Flows are equally distributed across production lines, and each flow inherits the average \gls{SNR} associated with its corresponding line. Production lines are placed at different distances from the \gls{BS}, while traffic flows are defined using a cyclic assignment over representative arrival rates and delay requirements from Table~\ref{tab:ExpSetup}.

For $|\mathcal{F}| \leq 18$, the execution time increases with $|\mathcal{F}|$ for all considered numbers of production lines, ranging from a few tens of seconds up to approximately $260$~s. This increase is mainly driven by the higher computational cost of individual iterations. As discussed in Section~\ref{sec:heuristic}, each iteration in both Phase~A and Phase~B requires a full recomputation of per-flow \gls{SNC}-based delay bounds, whose cost scales linearly with $|\mathcal{F}|$ (and with $|\mathcal{S}|$ in Phase~B), leading to longer per-iteration execution times as the number of flows grows. At the same time, the total number of iterations decreases with $|\mathcal{F}|$. As the same pool of $N_{\mathrm{cell}}^{\mathrm{RB}}$ \glspl{RB} is shared among more flows, the number of admissible \gls{RB} reallocations or removals becomes more limited, causing the heuristic to converge in fewer iterations. Despite this faster convergence, the increase in per-iteration cost dominates, resulting in higher overall execution times.

For a fixed number of flows, configurations with more production lines generally exhibit lower execution times. In these cases, the increased heterogeneity in average \gls{SNR} across production lines reduces the feasible region, causing Phase~B to terminate after fewer tightening iterations. The highest execution times are observed for $|\mathcal{F}| = 24$, where the heuristic remains close to feasibility and both Phase~A and Phase~B are executed for a non-negligible number of iterations, leading to a peak execution time close to $300$~s. In contrast, when radio resources are clearly insufficient to further reduce the objective function, infeasibility is detected early, Phase~A terminates after very few iterations with $\zeta > 1$, and Phase~B is skipped. This explains the markedly lower execution times observed for some configurations with $|\mathcal{F}| = 24$ and for all cases with $|\mathcal{F}| = 30$.

The execution times range from a few tens of seconds to several hundreds of seconds. Given that average traffic statistics evolve on time scales of minutes, these results support the feasibility of window-based network slice planning at the Non-\gls{RT} \gls{RIC} in the \gls{O-RAN} architecture.

\section{Conclusions and Future Works}\label{sec:conclusions}
This paper evaluated four \gls{RAN} slicing deployment options for \gls{5G}-based Industry~4.0 using a common \gls{SNC}-based slice planning framework. Within the considered \gls{RAN}-level analytical model, the results show that per-flow slicing (DO-\#2) is the only evaluated deployment option that consistently meets \gls{uRLLC} latency targets under limited spectrum availability, whereas per-line (DO-\#1), shared (DO-\#3), and hybrid (DO-\#4) deployments trade isolation for improved aggregation efficiency. The execution-time and scalability results indicate that the proposed planner operates at Non-\gls{RT} time scales, with execution times ranging from sub-second to several tens of seconds, supporting window-based slice planning in industrial \gls{O-RAN} deployments. Future work will extend the framework towards joint radio and \gls{TSN}-aware traffic coordination for tightly synchronized industrial automation systems.

\section*{Acknowledgment}
This work is part of the project PID2022-137329OB-C43 funded by MICIU/AEI/10.13039/501100011033 and by FEDER, EU, and also part of the project C-ING-306-UGR23 funded by Consejería de Universidad, Investigación e Innovación and by ERDF Andalusia Program 2021-2027.

\bibliographystyle{ieeetr}
\bibliography{references}

\end{document}

%% file: acronyms.tex
\newacronym{3GPP}{3GPP}{3rd Generation Partnership Project}

\newacronym{5G}{5G}{5th Generation}
\newacronym{5G-ACIA}{5G-ACIA}{5G Alliance for Connected Industries and Automation}
\newacronym{5G-NR}{5G-NR}{5G New Radio}
\newacronym{6G}{6G}{Sixth Generation}

\newacronym{AMC}{AMC}{Adaptive Modulation and Coding}
\newacronym{AC}{AC}{Admission Control}
\newacronym{AGV}{AGV}{Automated Guided Vehicle}
\newacronym{AR}{AR}{Augmented Reality}
\newacronym{ATS}{ATS}{Asynchronous Traffic Shaping}

\newacronym{BLER}{BLER}{Block Error Rate}
\newacronym{BWP}{BWP}{Bandwidth Part}
\newacronym{BS}{BS}{Base Station}
\newacronym{BSS}{BSS}{Business Support System}

\newacronym{CDF}{CDF}{Cumulative Distribution Function}
\newacronym{CCDF}{CCDF}{Complementary Cumulative Distribution Function}
\newacronym{CDMA}{CDMA}{Code Division Multiple Access}
\newacronym{CTMC}{CTMC}{Continuos-Time Markov Chain}
\newacronym{CSI}{CSI}{Channel State Information}
\newacronym{CP}{CP}{Control Plane}
\newacronym{CQI}{CQI}{Channel Quality Indicator}
\newacronym{CU}{CU}{Centralized Unit}

\newacronym{DL}{DL}{downlink}
\newacronym{DNC}{DNC}{Deterministic Network Calculus}
\newacronym{DRP}{DRP}{Dynamic Resource Provisioning}
\newacronym{DRL}{DRL}{Deep Reinforcement Learnning}
\newacronym{DU}{DU}{Distributed Unit}
\newacronym{DRB}{DRB}{Data Radio Bearer}

\newacronym{eMBB}{eMBB}{enhanced Mobile Broadband}
\newacronym{ETSI}{ETSI}{European Telecommunication Standards Institute}
\newacronym{EBB}{EBB}{Exponentially Bounded Burstiness}
\newacronym{EBF}{EBF}{Exponentially Bounded Fluctuation}
\newacronym{E2E}{E2E}{End-to-End}
\newacronym{EDF}{EDF}{Earliest Deadline First}
\newacronym{EM}{EM}{Expectation-Maximization}

\newacronym{FCFS}{FCFS}{First-come First-served}
\newacronym{FIFO}{FIFO}{First In First Out}

\newacronym{GBR}{GBR}{Guaranteed Bit Rate}
\newacronym{GMM}{GMM}{Gaussian Mixture Model}
\newacronym{GSMA}{GSMA}{Global System for Mobile Communications Association}
\newacronym{GST}{GST}{Generic Network Slice Template}
\newacronym{gNB}{gNB}{Next generation NodeB}

\newacronym{HDR}{HDR}{High Data Rate}

\newacronym{ITU}{ITU}{International Telecommunication Union}
\newacronym{IoT}{IoT}{Internet of Things}
\newacronym{ILP}{ILP}{Integer Linear Programming}
\newacronym{ICIC}{ICIC}{Inter-Cell Interference Cancellation}
\newacronym{IIOT}{IIoT}{Industrial Internet of Things}

\newacronym{LA}{LA}{Link Adaptation}
\newacronym{LOS}{LoS}{Line-of-Sight}
\newacronym{LSTM}{LSTM}{Long Short-Term Memory}
\newacronym{LTE}{LTE}{Long Term Evolution}

\newacronym{MAC}{MAC}{Medium Access Control}
\newacronym{MEC}{MEC}{Multi-access Edge Computing}
\newacronym{MCS}{MCS}{Modulation and Coding Scheme}
\newacronym{MDN}{MDN}{Mixture Density Network}
\newacronym{MGF}{MGF}{Moment Generating Function}
\newacronym{MIMO}{MIMO}{Multiple Input Multiple Output}
\newacronym{MISO}{MISO}{Multiple Input Single Output}
\newacronym{MRC}{MRC}{Maximal Ratio Combining}
\newacronym{ML}{ML}{Machine Learning}
\newacronym{MNO}{MNO}{Mobile Network Operator}
\newacronym{mMTC}{mMTC}{Machine Type Communication}
\newacronym{MSE}{MSE}{Mean Squared Error}
\newacronym{mURLLC}{mURLLC}{massive ultra-Reliable Low Latency Communication}

\newacronym{NE}{NE}{Nash Equilibrium}
\newacronym{NEST}{NEST}{Network Slice Type}
\newacronym{NIP}{NIP}{Non-linear Integer Programming}
\newacronym{NFMF}{NFMF}{Network Function Management Function}
\newacronym{NFV}{NFV}{Network Function Virtualization}
\newacronym{NG-RAN}{NG-RAN}{Next Generation - RAN}
\newacronym{NLOS}{NLoS}{Non-Line-of-Sight}
\newacronym{NN}{NN}{Neural Network}
\newacronym{NSO}{NSO}{Network Slice Orchestrator}
\newacronym{NSMF}{NSMF}{Network Slice Management Function}
\newacronym{NSSMF}{NSSMF}{Network Slice Subnet Management Function}
\newacronym{NR}{NR}{New Radio}

\newacronym{OFDMA}{OFDMA}{Orthogonal Frequency-Division Multiple Access}
\newacronym{O-RAN}{O-RAN}{Open Radio Access Network}
\newacronym{PDF}{PDF}{Probability Density Function}
\newacronym{PMF}{PMF}{Probability Mass Function}
\newacronym{PRB}{PRB}{Physical Resource Block}
\newacronym{P-NEST}{P-NEST}{private NEST}
\newacronym{PDU}{PDU}{Protocol Data Unit}
\newacronym{PLC}{PLC}{Programmable Logic Controller}

\newacronym{QoS}{QoS}{Quality of Service}
\newacronym{QFI}{QFI}{QoS Flow Identifier}

\newacronym{RAN}{RAN}{Radio Access Network}
\newacronym{RB}{RB}{Resource Block}
\newacronym{RBG}{RBG}{Resource Block Group}
\newacronym{RIC}{RIC}{RAN Intelligent Controller}
\newacronym{RIS}{RIS}{Reconfigurable Intelligent Surfaces}
\newacronym{RRM}{RRM}{Radio Resource Management}
\newacronym{RSMA}{RSMA}{Rate Splitting Multiple Access}
\newacronym{RSRP}{RSRP}{Received Signal Received Power}
\newacronym{RSRQ}{RSRQ}{Received Signal Received Quality}
\newacronym{RSSI}{RSSI}{Received Signal Strength Indication}
\newacronym{RT}{RT}{Real Time}
\newacronym{RU}{RU}{Radio Unit}

\newacronym{SDO}{SDO}{Standards Developing Organization}
\newacronym{SINR}{SINR}{Signal-to-Interference-plus-Noise Ratio}
\newacronym{SLA}{SLA}{Service Level Agreement}
\newacronym{SNC}{SNC}{Stochastic Network Calculus}
\newacronym{SNR}{SNR}{Signal-to-Noise Ratio}
\newacronym{S-NEST}{S-NEST}{standardized NEST}
\newacronym{SIMO}{SIMO}{Single-Input Multiple-Output}

\newacronym{TTI}{TTI}{Transmission Time Interval}
\newacronym{TSN}{TSN}{Time-Sensitive Networking}
\newacronym{TAS}{TAS}{Time-Aware Shaper}

\newacronym{UE}{UE}{User Equipment}
\newacronym{UL}{UL}{Uplink}
\newacronym{UP}{UP}{User Plane}
\newacronym{uRLLC}{uRLLC}{ultra-Reliable Low Latency Communication}

\newacronym{V2X}{V2X}{Vehicle-to-Everything}
\newacronym{VBR}{VBR}{Variable Bit Rate}
\newacronym{VR}{VR}{Virtual Reality}
\newacronym{vRAN}{vRAN}{virtualized RAN}
\newacronym{vBS}{vBS}{virtualized Base Station}
\newacronym{VS}{VS}{Validation Scenario}

\newacronym{WiMAX}{WiMAX}{Worldwide Interoperability for Microwave Access}
\newacronym{WCDMA}{WCDMA}{Wideband \gls{CDMA}}
\newacronym{WRR}{WRR}{Weighted Round Robin}